\documentclass[%
aps, prl,
floatfix,
preprintnumbers,
reprint,
superscriptaddress,
amsmath, amssymb,
longbibliography,
]{revtex4-2}

\usepackage{soul}

\usepackage[utf8]{inputenc}
\usepackage{graphicx}
\usepackage{bm}
\usepackage[breaklinks=true]{hyperref}
\hypersetup{bookmarksnumbered, pdfpagemode=UseOutlines, 
	pdfauthor={K.\ S.\ Das}, 
	pdftitle={Modulation of magnon spin transport in a magnetic gate transistor}, 
	pdfdisplaydoctitle, 
	colorlinks=true, citecolor=blue, filecolor=blue, linkcolor=blue, urlcolor=blue}

\begin{document}
	
	\title{Modulation of magnon spin transport in a magnetic gate transistor}
	
	\author{K.\ S.\ \surname{Das}}
	\email[e-mail: ]{K.S.Das@rug.nl}
	\affiliation{Physics of Nanodevices, Zernike Institute for Advanced Materials, University of Groningen, 9747 AG Groningen, The Netherlands}
	\author{F. Feringa}
	\affiliation{Physics of Nanodevices, Zernike Institute for Advanced Materials, University of Groningen, 9747 AG Groningen, The Netherlands}
	\author{M. Middelkamp}
	\affiliation{Physics of Nanodevices, Zernike Institute for Advanced Materials, University of Groningen, 9747 AG Groningen, The Netherlands}
	\author{B.\ J.\ \surname{van Wees}}
	\email[e-mail: ]{B.J.van.Wees@rug.nl}
	\affiliation{Physics of Nanodevices, Zernike Institute for Advanced Materials, University of Groningen, 9747 AG Groningen, The Netherlands}
	\author{I.\ J.\ Vera-Marun}
	\email[e-mail: ]{ivan.veramarun@manchester.ac.uk}
	\affiliation{School of Physics and Astronomy, University of Manchester, Manchester M13 9PL, United Kingdom}
	
	\date{\today}
	
	\begin{abstract}
		We demonstrate a modulation of up to $18\%$ in the magnon spin transport in a magnetic insulator ($\text{Y}_3\text{Fe}_5\text{O}_{12}$, YIG) using a common ferromagnetic metal (permalloy, Py) as a magnetic control gate. A Py electrode, placed between two Pt injector and detector electrodes, acts as a magnetic gate in our prototypical magnon transistor device. By manipulating the magnetization direction of Py with respect to that of YIG, the transmission of magnons through the Py$\mid$YIG interface can be controlled, resulting in a modulation of the non-equilibrium magnon density in the YIG channel between the Pt injector and detector electrodes. This study opens up the possibility of using the magnetic gating effect for magnon-based spin logic applications.     
	\end{abstract}
	
	
	
	\maketitle
	Magnon-based spintronic devices are alternatives for charge-based electronics  \cite{chumak_magnon_2015,jungwirth_antiferromagnetic_2016,baltz_antiferromagnetic_2018}. Information, in the form of spin waves or magnons, can be transmitted over a long distance in magnetic insulators \cite{cornelissen_long-distance_2015}, without the necessity of accompanying electron transport. Thus, magnon-based devices can be used as a new type of interconnects in spintronic circuitry. Additionally, the modulation of magnon spin transport would also enable the use of such devices for logic operations \cite{chumak_magnon_2014}. This has led to a recent surge in experiments exploring the control of magnon transport via magnon-valves \cite{wu_magnon_2018,guo_magnon_2018,cramer2018magnon} and in the magnon transistor geometry \cite{cornelissen_spin-current-controlled_2018}.\\
	In order to implement magnonic logic devices, two distinctive features are important: control of the magnon spin transport and storage of information in a memory device. The presence of both functionalities in a single device is still missing in the previous mentioned magnon-valves and magnon transistor. In this work we present a solution using a novel geometry for a magnon transistor, making use of a magnetic gating effect, which also has the potential to implement memory and therefore be programmable. \\
		\begin{figure}[tbp]
		\includegraphics*[angle=0, trim=0mm 0mm 0mm 0mm, width=90mm]{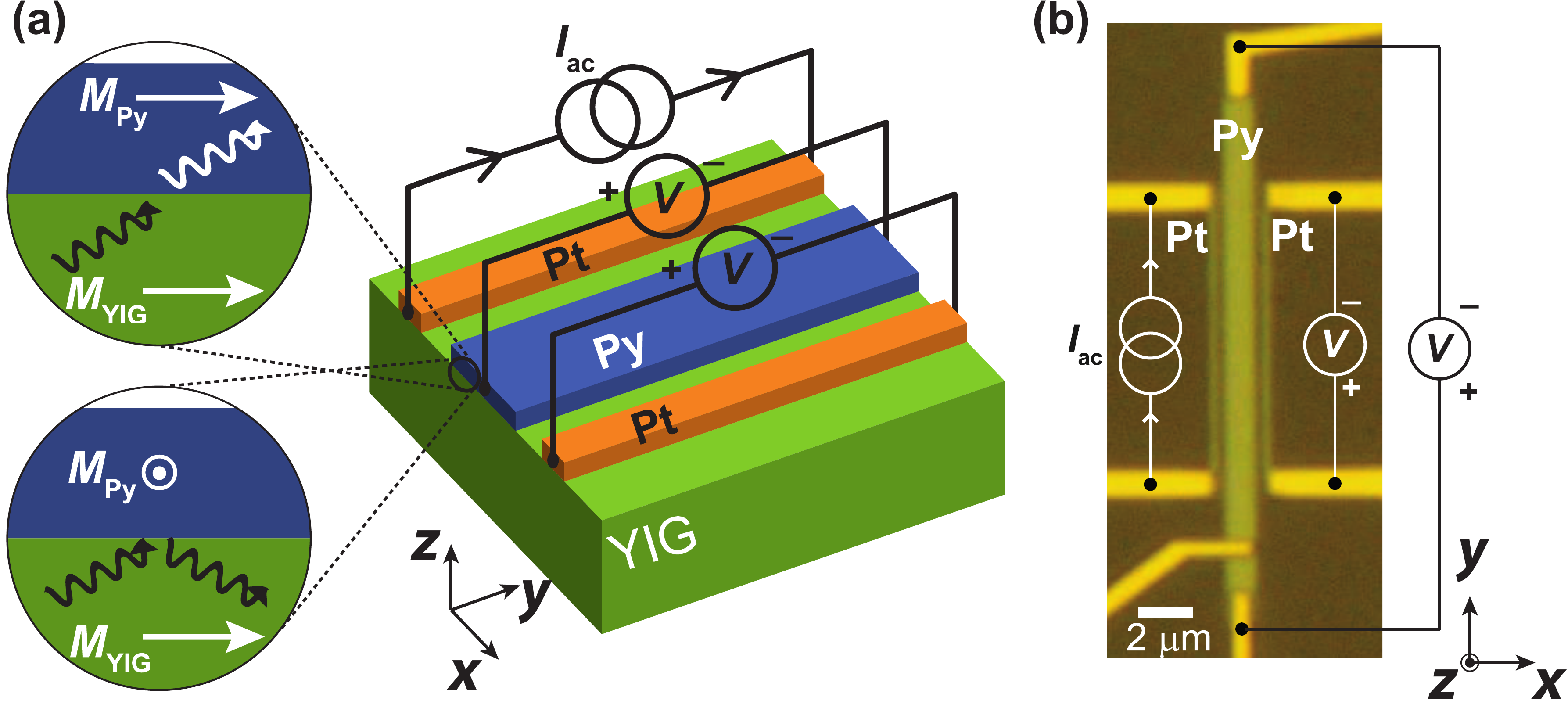}
		\caption{
			\label{fig:9-1}
			\textbf{(a)} Schematic illustration of the device geometry. The magnon-valve effect is depicted in the insets, where the transmission of magnons across the Py$\mid$YIG interface is dependent on the relative orientation of the Py ($\textbf{\textit{M}}_\text{Py}$) and the YIG ($\textbf{\textit{M}}_\text{YIG}$) magnetizations. \textbf{(b)} An optical image of the device is shown, along with the electrical connections for the non-local magnon transport experiment. The centre-to-centre distance between the Pt injector and the Pt detector is $2~\mu\text{m}$ for all the devices. 
		}
	\end{figure}	
	The ferromagnetic metal permalloy (Py) is used in a proof-of-concept device geometry for demonstrating the modulation of magnon spin transport in a magnetic insulator (YIG) via the magnetic gating effect. Exchange (thermal) magnons are injected using a Pt electrode via the spin Hall effect (SHE), resulting in a non-equilibrium magnon accumulation in the YIG film \cite{cornelissen_long-distance_2015,cornelissen_magnon_2016}. A second Pt electrode is used to electrically detect the non-equilibrium magnons via the inverse spin Hall effect (ISHE). A middle Py strip is placed between the Pt injector and detector electrodes for manipulating the magnon transport in the YIG channel via the magnetic gating effect, schematically depicted in Fig.~\ref{fig:9-1}(a). When a magnon arrives at the YIG$\mid$Py interface, three magnon absorption mechanisms are possible: spin-flip scattering at the interface generating a spin accumulation in the Py, spin transfer torque and magnon to magnon transmission across the YIG$\mid$Py interface.  The spin transfer torque is maximum when the Py and the YIG magnetizations, $\textbf{\textit{M}}_\text{Py}$ and $\textbf{\textit{M}}_\text{YIG}$, are oriented perpendicular to each other.  When $\textbf{\textit{M}}_\text{Py}$ and $\textbf{\textit{M}}_\text{YIG}$ are oriented parallel to each other, the transmission of magnons from the YIG film into the Py strip and spin-flip scattering at the interface is maximized. Considering a shorter magnon mean free path in Py as compared to YIG \cite{boona_magnon_2014,chavez-angel_reconstruction_2017}, the transmission of magnons into the Py would lead to a decrease in the non-equilibrium magnon density in the YIG channel. This will result in the modulation of the non-local magnon spin signal measured by the Pt detector as a function of the relative orientation between $\textbf{\textit{M}}_\text{Py}$ and $\textbf{\textit{M}}_\text{YIG}$. Therefore, if spin transfer torque is a dominant process, a decrease of the magnon current is expected when $\textbf{\textit{M}}_\text{Py}$ and $\textbf{\textit{M}}_\text{YIG}$ are oriented perpendicular. When spin-flip scattering and magnon to magnon transmission are dominant a decrease of the magnon current is expected for parallel alignment of $\textbf{\textit{M}}_\text{Py}$ and $\textbf{\textit{M}}_\text{YIG}$. The similar geometry as in Ref.~\onlinecite{cornelissen_spin-current-controlled_2018} has been used, but the modulation mechanism is completely different in nature. The modulation in Ref.~\onlinecite{cornelissen_spin-current-controlled_2018} is achieved by creating a non-equilibrium magnon density in the YIG via the electrically-driven Pt modulator, whereas in this work the magnons in the YIG channel are in equilibrium with the modulator. We demonstrate that a modulation of up to $18\%$ can be achieved in our devices, which is more than an order of magnitude higher than that reported in Ref.~\onlinecite{cornelissen_spin-current-controlled_2018} for the same YIG film thickness (210~nm), using a Pt modulator. 
	
	\begin{figure*}[tbp]
		\includegraphics*[angle=0, trim=0mm 0mm 0mm 0mm, width=170mm]{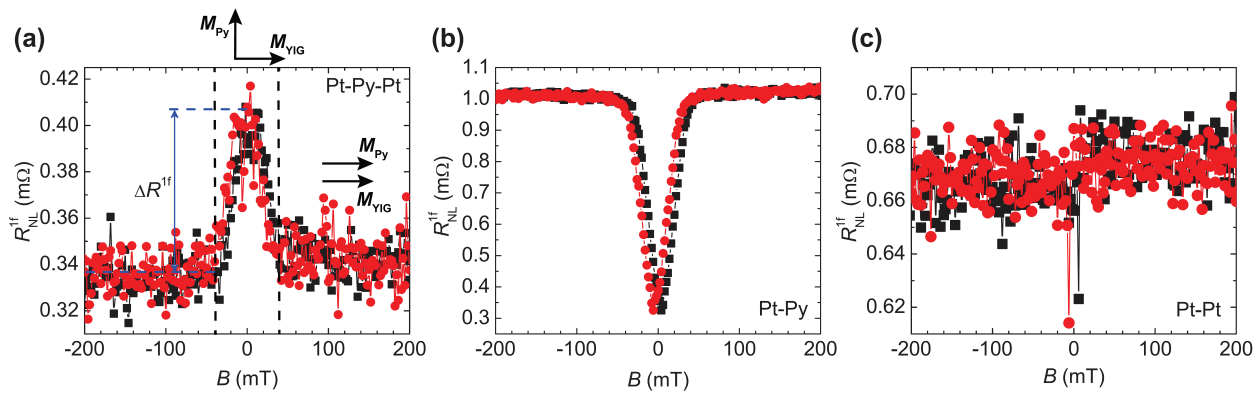}
		\caption{
			\label{fig:9-2}
			A magnetic field ($B$) is swept by increasing the field (trace, black) and decreasing the field (retrace, red) along the $x$-axis and the first harmonic response of the non-local magnon spin signal ($R_\text{NL}^\text{1f}$) is measured by \textbf{(a)} Pt as injector and detector in a device with a 900~nm wide Py middle strip (Pt-Py-Pt), \textbf{(b)} Pt as injector and the 900~nm wide middle Py strip as detector (Pt-Py), and \textbf{(c)} Pt as injector and detector in a reference device without any middle Py strip (Pt-Pt). The arrows in \textbf{(a)} indicate the relative orientation of the magnetizations of Py and YIG. Py has a thickness of 9 nm.}
	\end{figure*}
	Three batches of devices were fabricated using electron beam lithography on 210$\,$nm thick YIG (111) films, grown by liquid-phase epitaxy on GGG ($\text{Gd}_3\text{Ga}_5\text{O}_{12}$) substrates. 7~nm thick Pt strips, with widths of 200~nm, were d.c.\ sputtered on YIG as the injector and detector electrodes. The dimensions of the middle Py strip, also fabricated by d.c.\ sputtering, were varied among the different batches of devices. In the first two batches we varied the Py width (300, 500, 600 and 900 nm) while keeping a constant thickness of 9 nm, whereas in the third batch we varied the Py thickness (9, 15 and 38 nm) while keeping a constant width of 900 nm. A Pt-Pt device without any middle Py strip was fabricated as a reference device. The centre-to-centre distance between the Pt injector and detector electrodes was kept constant at $2~\mu\text{m}$ for all the devices. An optical image of a device with a 900~nm wide middle Py strip is shown in Fig.~\ref{fig:9-1}(b), along with the electrical connections. An alternating current ($I$), with an rms amplitude of 400$\,$$\mu$A and frequency of 11$\,$Hz, was sourced through the Pt injector (left). The first (1f) and the second harmonic (2f) responses of the non-local voltage ($V$), correspond to the electrically injected (via the SHE) and the thermally-injected (via the spin Seebeck effect driven by Joule heating at the injector) magnons. Both responses were measured simultaneously across the Pt detector and the middle Py strip, by a phase-sensitive lock-in detection technique. The non-local magnon spin signal is defined as $R_\text{NL}^\text{1f}=V^\text{1f}/I$ for the electrically injected magnons, and $R_\text{NL}^\text{2f}=V^\text{2f}/I^2$ for the thermally injected magnons. All the measurements were carried out at room temperature.\\
	\begin{figure}[b]
		\includegraphics*[angle=0, trim=0mm 0mm 0mm 0mm, width=85mm]{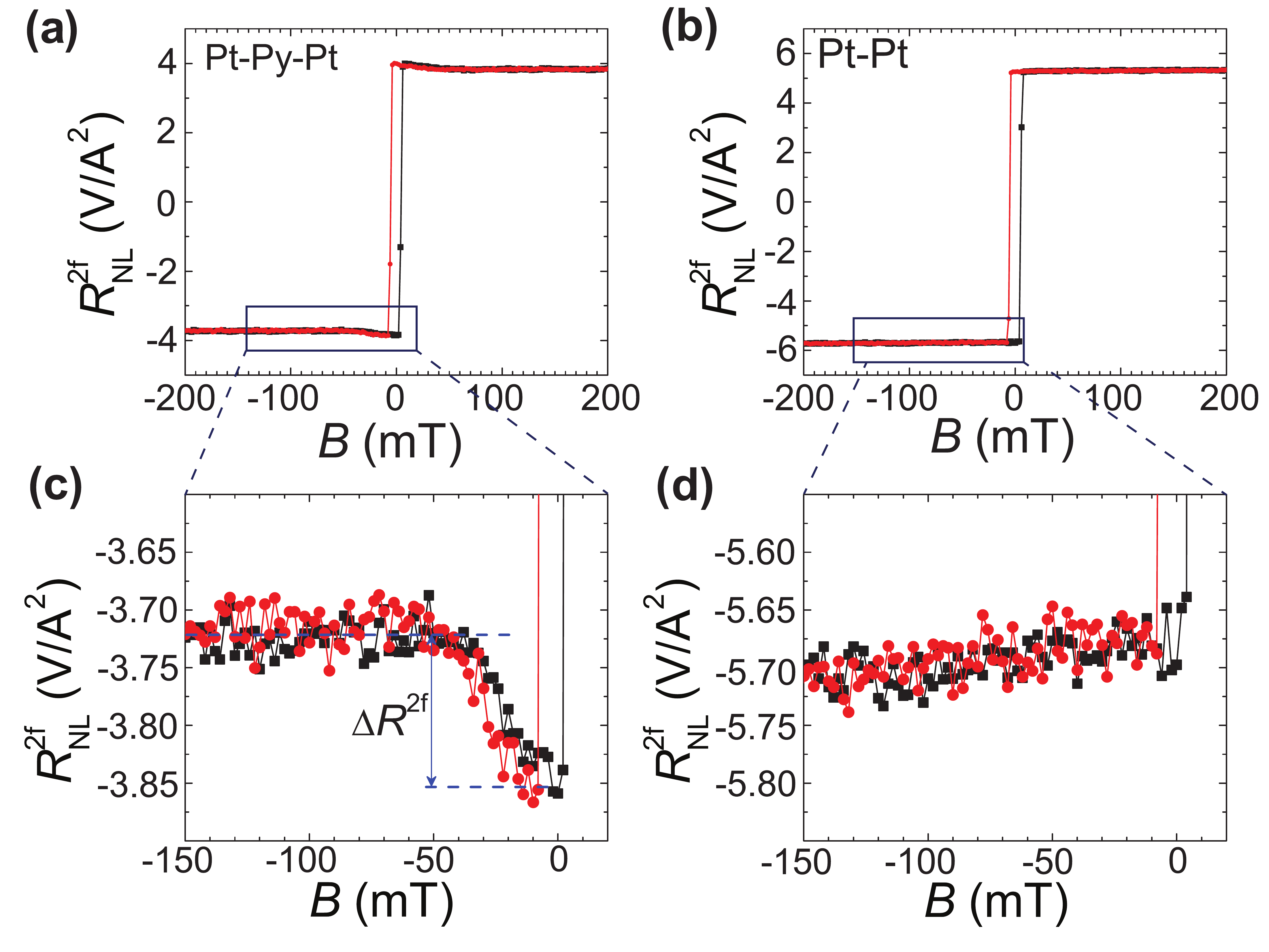}
		\caption{
			\label{fig:9-3}
			Second harmonic response of the non-local magnon spin signal ($R_\text{NL}^\text{2f}$), measured as a function of $B$, by \textbf{(a)} the Pt detector in a device with a 900~nm wide Py middle strip, and \textbf{(b)} in a reference device without a middle Py strip. Magnified regions from the graphs in \textbf{(a)} and \textbf{(b)} are shown in \textbf{(c)} and \textbf{(d)}, respectively, demonstrating the effect of the middle Py strip on $R_\text{NL}^\text{2f}$. The data shown in black and red represent the trace and retrace directions, respectively.  
		}
	\end{figure} 
	An external magnetic field ($B$) was swept along the hard axis ($x$-axis) direction of the magnetic gate and the corresponding $R_\text{NL}^\text{1f}$ measured, as shown in Figs.~\ref{fig:9-2}(a-c). A modulation in $R_\text{NL}^\text{1f}$, measured by the Pt detector in the devices with a middle Py, was observed [see Fig.~\ref{fig:9-2}(a)]. The maximum value of $R_\text{NL}^\text{1f}$ occurs at $B=0$, when $\textbf{\textit{M}}_\text{Py}$ is oriented along the easy axis of the magnetic gate ($y$-axis), perpendicular to $\textbf{\textit{M}}_\text{YIG}$. Note that due to a low coercive field of our YIG film ($<1$~mT) \cite{vlietstra_spin-hall_2013}, $\textbf{\textit{M}}_\text{YIG}$ is essentially always oriented along the $x$-axis in our measurements. Besides, any possible interfacial exchange interaction between Py and YIG doesn't play a significant role \cite{das_efficient_2018,das_spin_2017} and therefore $\textbf{\textit{M}}_\text{YIG}$ and $\textbf{\textit{M}}_\text{Py}$ can move freely with respect to each other. By changing the magnitude of $B$, $R_\text{NL}^\text{1f}$ was modulated, reaching a minimum value at $|B|\approx50$~mT, corresponding to the tilting of $\textbf{\textit{M}}_\text{Py}$ along the in-plane hard axis direction ($x$-axis) of the magnetic gate. For $|B|\geq50$~mT, when $\textbf{\textit{M}}_\text{Py}$ and $\textbf{\textit{M}}_\text{YIG}$ are aligned parallel to each other, $R_\text{NL}^\text{1f}$ decreases to its minimum value, corresponding to a modulation ($\Delta R^\text{1f}$) of about $18\%$. Therefore, the electrically injected magnons from the Pt injector reaching the Pt detector decrease by $18\%$ by reorienting the magnetization direction of the Py gate electrode.

	$R_\text{NL}^\text{1f}$ measured across the middle Py strip is shown in Fig.~\ref{fig:9-2}(b). The detection of non-local magnon transport at the Py strip occurs via a combination of ISHE and the inverse anomalous spin Hall effect (IASHE), resulting in a detection efficiency that depends on the orientation of $\textbf{\textit{M}}_\text{Py}$ and leading to a line shape consistent with previous reports \cite{das_spin_2017,das_efficient_2018}. A modulation of more than $210\%$ in the $R_\text{NL}^\text{1f}$ measured by the Py detector is observed, which occurs due to the detection mechanism being dominated only by ISHE at low $B$ and evolving into being composed by both IASHE and ISHE at high $B$. Note that this modulation in the magnon detection efficiency, within the Py electrode, is one order of magnitude larger, and of a different nature, than the $18\%$ modulation seen in Fig.~\ref{fig:9-2}(a), which is due to the modulation of magnon current in the YIG between the Pt injector and detector.

	The non-local signal measured in a reference Pt-Pt device, without any middle Py strip, is shown in Fig.~\ref{fig:9-2}(c). $R_\text{NL}^\text{1f}$ was found to be constant in this reference device, which evidences the role of the middle Py strip in the modulation of $R_\text{NL}^\text{1f}$ in the non-reference devices, as shown in Fig.~\ref{fig:9-2}(a). Therefore, we can modulate the magnon current reaching the Pt detector using the Py gate, due to a modulation of the magnon absorption in the Py.
	\begin{figure*}[tbp]
		\includegraphics*[angle=0, trim=0mm 0mm 0mm 0mm, width=170mm]{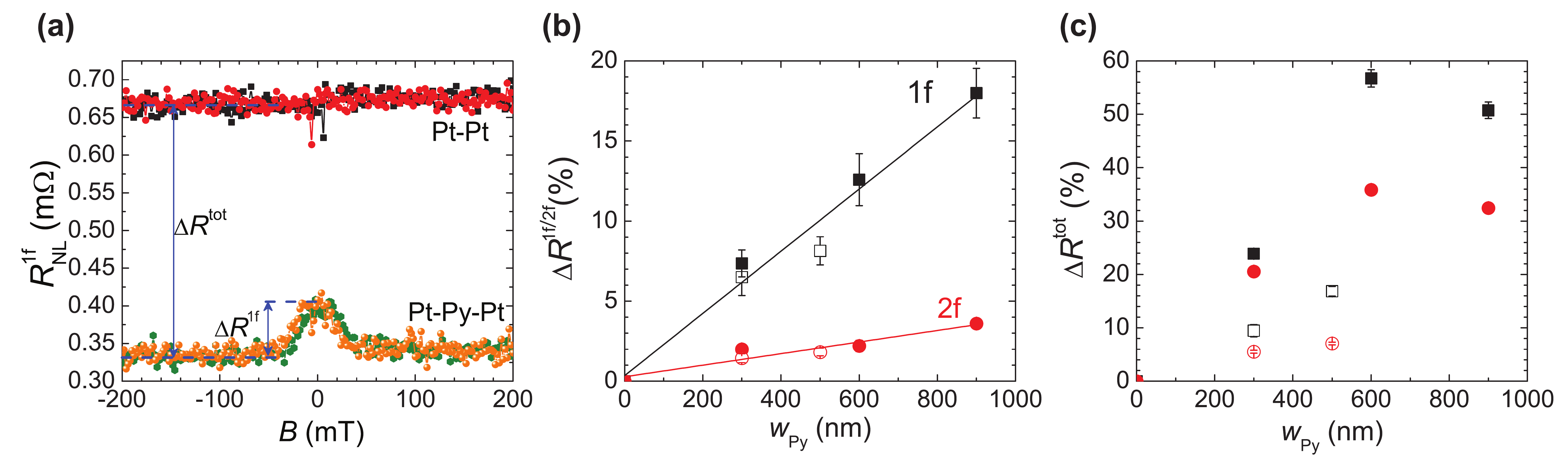}
		\caption{
			\label{fig:9-4}
			\textbf{(a)} $R_\text{NL}^\text{1f}$, measured by the Pt detector in the reference device without the middle Py strip [black(trace)-red(retrace)] and in the device with a 900~nm wide middle Py strip [green(trace)-orange(retrace)] are plotted together, illustrating the relative modulation ($\Delta R^\text{1f(2f)}$) and the total modulation ($\Delta R^\text{tot}$) in the non-local magnon spin signal. $\Delta R^\text{1f(2f)}$ \textbf{(b)} and ($\Delta R^\text{tot}$) \textbf{(c)} are plotted as a function of the middle Py width ($w_\text{Py}$). The black squares and the red circles represent the modulations in the first and the second harmonic response of the non-local signal, respectively, while the open and the filled symbols correspond to devices from two different batches. The linear dependence of $\Delta R^\text{1f(2f)}$ on $w_\text{Py}$ is evident from the linear fits (solid lines) to the data in \textbf{(b)}. $\Delta R^\text{1f/2f}(\%)$ and $\Delta R^\text{\text{tot}}(\%)$ are defined as $\Delta R^\text{1f/2f}(\%) = \Delta R^\text{1f/2f}/R^\text{1f/2f}(|B|\geq50\text{~mT})$ and $\Delta R^\text{\text{tot}}(\%) =\Delta R^\text{\text{tot}}/R^\text{Pt-Pt}$.
		}
	\end{figure*}\\The second harmonic response of the non-local magnon signal ($R_\text{NL}^\text{2f}$) was also measured by sweeping $B$ along the $x$-axis, as shown in Figs.~\ref{fig:9-3}(a-d). The magnetic gating effect of the middle Py strip also led to a modulation in $R_\text{NL}^\text{2f}$ measured by the Pt detector, as depicted in Figs.~\ref{fig:9-3}(a) and (c). However, the modulation in $R_\text{NL}^\text{2f}$ was found to be $\Delta R^\text{2f}\approx 3.6\%$, which is 5 times smaller than that of $R^\text{1f}$. Note that the second harmonic response is related to the non-equilibrium magnons which are generated via the spin Seebeck effect (SSE) in YIG \cite{uchida_spin_2010,cornelissen_long-distance_2015,cornelissen_magnon_2016}, driven by the thermal gradient created by the Pt injector due to Joule heating. The spacing between injector and detector is large and it corresponds to a magnon accumulation at the detector, confirmed by the correct sign of the second harmonic response \cite{shan2016influence}.  The transmission of these thermally generated magnons into the middle Py strip also depends on the relative orientation of $\textbf{\textit{M}}_\text{Py}$ and $\textbf{\textit{M}}_\text{YIG}$, resulting in the modulation of $R_\text{NL}^\text{2f}$ [Figs.~\ref{fig:9-3}(a) and (c)]. To rationalize the smaller modulation in the 2f signal, as compared to the 1f signal, we have to look at the magnons generated by SSE. The temperature gradient extends through the YIG sample and therefore, thermally-generated magnons are not only generated close to the injector, but also in the region between the injector and detector. Magnons accumulate at the bottom of the YIG film and then diffuse towards the detector \cite{PhysRevB.96.184427}. Therefore, less thermally-generated magnons cross the modulator at the interface while diffusing towards the detector, which results in a relatively smaller modulation of the second harmonic signal in comparison to the first harmonic signal.  Note that the total magnitude of $R_\text{NL}^\text{2f}$ is reduced compared to the case of having no middle Py strip in the reference Pt-Pt device, as shown in Figs.~\ref{fig:9-3}(b) and (d). In this reference device, there is no modulation in $R_\text{NL}^\text{2f}$ with $B$.
	
	Furthermore, we study the dependence of the modulation of the non-local magnon spin signals on the width of the middle Py gate ($w_\text{Py}$). We define a relative modulation for the first ($\Delta R^{\text{1f}}$) and second harmonic ($\Delta R^{\text{2f}}$) signals and a total modulation ($\Delta R^{tot}$) of the spin signal, as depicted in Fig.~\ref{fig:9-4}(a). $\Delta R^{\text{1f(2f)}}$ gives the modulation only due to the magnetization orientation dependent magnetic gating effect, whereas, $\Delta R^{tot}$ gives the total modulation of the spin signal compared to the reference device (without any middle Py strip). We find a linear dependence of $\Delta R^{\text{1f(2f)}}$ on $w_\text{Py}$ for both first (second) 1f (2f) harmonic response of the spin signal, as shown in Fig.~\ref{fig:9-4}(b). Also, the variation in $\Delta R^{\text{1f(2f)}}$ between two different batches of devices (depicted as open and filled symbols) is very small, demonstrating the reproducibility of the magnetic gating effect. In the case of the total modulation $\Delta R^{tot}$ although it exhibits an increasing trend with $w_\text{Py}$, the results are dominated by a batch-to-batch variability. We attribute this variability to the difference in transparencies at Pt$\mid$YIG and Py$\mid$YIG interfaces amongst the different batches of devices. On the other hand, the relative modulation $\Delta R^{\text{1f(2f)}}$ filters out any geometrical or interfacial variation and exhibits a clear linear scaling with $w_\text{Py}$. Given the long magnon relaxation length in YIG ($\approx 10~\mu\text{m}$) \cite{cornelissen_long-distance_2015}, we expect the decay of the magnon chemical potential between the Pt injector and detector electrodes to be slow for a separation of $2~\mu\text{m}$ in our devices. Therefore, the linear scaling with $w_\text{Py}$ further supports the magnetization orientation dependent magnon absorption into the middle Py gate.
	
		\begin{figure}[b]
		\includegraphics*[angle=0, trim=0mm 0mm 0mm 0mm, width=85mm]{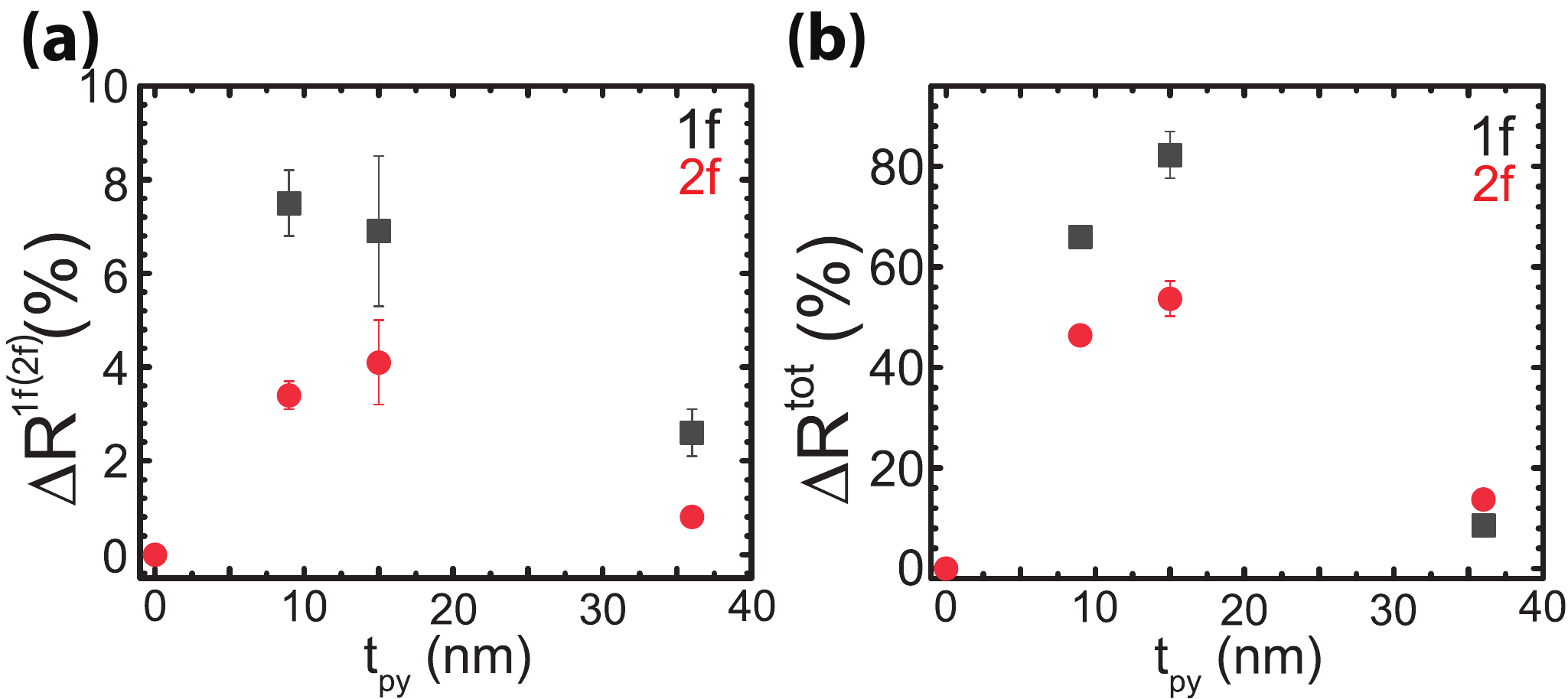}
		\caption{
			\label{fig:9-5}
			$\Delta R^\text{1f(2f)}$ \textbf{(a)} and $\Delta R^\text{tot}$ \textbf{(b)} are plotted as a function of the thickness of the middle Py strip ($t_\text{Py}$), which have a width of 900~nm. The black squares and the red circles represent the modulations in the first and the second harmonic response of the non-local signal, respectively.    
		}
		
	\end{figure} Finally, we study the dependence of the modulation of the non-local magnon spin signals on the thickness ($t_\text{Py}$) of the middle Py strip. $\Delta R^{\text{1f(2f)}}$ and $\Delta R^{tot}$ as a function of Py thickness is shown in Fig.~\ref{fig:9-5}(a) and (b). $\Delta R^{\text{1f(2f)}}$ and $\Delta R^{tot}$ show an increase for increasing Py thickness up to 15~nm, after which $\Delta R^{\text{1f(2f)}}$ and $\Delta R^{tot}$  show a decrease for increasing Py thickness. The devices with different Py thicknesses involve at least one additional lithography step compared to the first two batches of devices. This can significantly influence the Py$\mid$YIG and Pt$\mid$YIG interface, causing the difference in $\Delta R^{tot}$ and $\Delta R^{\text{1f(2f)}}$ between the first two and the third batch of devices. Besides, the Gilbert damping in Py is lower for thicker Py films \cite{zhao2016experimental}, meaning a longer magnon relaxation length and therefore the finite probability of magnons travelling into the Py and back into the YIG is larger for thicker Py films than for thinner Py films. This can explain the lower $\Delta R^{\text{1f(2f)}}$ modulation using a thicker Py gate.

	In this study, we have demonstrated efficient modulation of non-local magnon spin transport in a magnetic insulator using a magnetic gate in a proof-of-concept transistor device geometry. We achieve a modulation of up to an order of magnitude larger than a previously reported three-terminal magnon transistor \cite{cornelissen_spin-current-controlled_2018} with the same YIG thickness, where the spin transport was modulated by creating a non-equilibrium magnon density in the YIG channel via an electrically-driven Pt gate. In this work we show that the spin transport in the YIG channel is modulated in (or close to) equilibrium at the ferromagnetic metal$\mid$magnetic insulator interface, where the magnon transmission at the interface is controlled by manipulating the magnetization direction of a magnetic gate. A decrease in the magnon current is observed for parallel orientation of the magnetizations. Therefore, we conclude that spin-flip scattering and magnon to magnon transmission dominates over spin transfer torque at ferromagnetic metal$\mid$magnetic insulator interface. The origin of the magnetic gating effect is either an enhancement of spin-flip scattering or magnon to magnon transmission at the Py$\mid$YIG interface. We propose that such a magnetic gate can be used for future magnon transistor spin logic applications and memory applications embedded in the ferromagnetic gate, which can be used in programmable magnonics devices.
	\begin{acknowledgments}
	We acknowledge the technical support from J.\ G.\ Holstein, H.\ M.\ de Roosz, H.\ Adema T.\ Schouten and H. de Vries. We acknowledge the financial support of the Zernike Institute for Advanced Materials and the Future and Emerging Technologies (FET) programme within the Seventh Framework Programme for Research of the European Commission, under FET-Open Grant No.~618083 (CNTQC). This project is also financed by the NWO Spinoza prize awarded to Prof.\ B.\ J.\ van Wees by the NWO.		
	\end{acknowledgments}\bibliographystyle{apsrev4-2}
	%
	
\end{document}